# Ionization waves of arbitrary velocity driven by a flying focus

J.P. Palastro, D. Turnbull, S.-W. Bahk, R.K. Follett, J.L. Shaw, D. Haberberger,
J. Bromage, and D.H. Froula
*University of Rochester, Laboratory for Laser Energetics, Rochester NY, USA*

**Abstract**

A chirped laser pulse focused by a chromatic lens exhibits a dynamic, or "flying," focus in which the trajectory of the peak intensity decouples from the group velocity. In a medium, the flying focus can trigger an ionization front that follows this trajectory. By adjusting the chirp, the ionization front can be made to travel at an arbitrary velocity along the optical axis. We present analytical calculations and simulations describing the propagation of the flying focus pulse, the self-similar form of its intensity profile, and ionization wave formation. The ability to control the speed of the ionization wave and, in conjunction, mitigate plasma refraction has the potential to advance several laser-based applications, including Raman amplification, photon acceleration, high harmonic generation, and THz generation.

## I. Introduction

Several laser-plasma-based applications rely critically on the controlled formation and propagation of an ionization front. The speed of the ionization front can determine a phase matching condition, as in THz [1-3] or high harmonic generation [4-7]; the interaction distance, as in photon acceleration [8-10]; or the plasma conditions within an interaction region, as in Raman amplification [11-14]. Conventionally, a laser-produced ionization front is constrained to travel at the group velocity. This can severely limit the potential of the aforementioned applications.

Here we present a novel method for controlling the speed of an ionization front based on the recently demonstrated "flying focus" [15]. In its simplest implementation, the flying focus (FF) occurs when a chromatic lens focuses a chirped laser pulse [15,16]. In a region about the focus, the peak intensity of the laser pulse propagates at a speed that is decoupled from its group velocity. Figure (1) illustrates this schematically. The top row shows the chromatic focusing of a positively chirped pulse. The red-shifted frequencies lead the blue-shifted frequencies in time and focus closer to the lens. The resulting intensity peak travels from left to right, copropagating with the pulse. The bottom row shows the opposite case of a negatively chirped pulse. The blue-shifted frequencies lead in time, but focus further from the lens than the red-shifted frequencies. The intensity peak now travels from right to left, counter-propagating with respect to the pulse. More generally, the velocity of the intensity peak, or focal velocity, can take any value, depending on the focal length of the chromatic lens and the bandwidth and chirp of the pulse.

In a medium, the intensity peak of the FF can trigger an ionization front that travels at the focal velocity. This allows for an ionization wave of arbitrary velocity (IWAV) by adjusting the chirp. Because the velocity can take any value, the IWAV has the potential to enable a wide range of applications currently limited or precluded by an inability to phase-match at the group velocity of the laser pulse. Additionally, a FF pulse with negative focal velocity can create a contiguous IWAV relatively undisrupted by ionization refraction [17,18]. The simplicity of IWAV formation, requiring only a chirp and a chromatic lens, compares favorably to other schemes, for instance, using the intersection of two cylindrically focused femtosecond pulses [19].

We begin by reviewing the propagation characteristics of the flying focus, deriving expressions for the trajectory and velocity of the intensity peak. We show analytically that the FF pulse exhibits self-similar behavior, largely maintaining its spatiotemporal profile as it travels through the focal region. After discussing the FF propagation, we present self-consistent simulations of IWAV formation. The negative focal velocity pulses create a sharp ionization front by largely avoiding plasma refraction. For the parameters considered, the self-similar behavior persists in the presence of ionization. Finally, we show that modifications to the power spectrum of the FF pulse provide an additional avenue for tailoring the IWAV. As a specific example, we modify the power spectrum to create a plasma density profile modulated along the optical axis.

**II. Flying Focus Propagation Model**

To model the propagation of a flying focus pulse, we employ a "modified" paraxial wave equation (MPE). The MPE evolves the spatiotemporal envelope of a laser pulse and is equivalent to solving the paraxial wave equation for each frequency component within the envelope. The MPE captures two relevant effects not accounted for in a monochromatic paraxial equation: temporal delays due to path length differences within a pulse and the frequency dependence of diffraction. Both of these effects play a significant role in the propagation of a FF pulse.

We express the electric field of the pulse as a plane wave modulating an envelope: $\mathbf{E} = \frac{1}{2}\mathbf{E}_0(\mathbf{r},\xi,z)e^{i(k_0 z - \omega_0 t)} + \text{c.c.}$, where $\xi = ct - z$ is the moving frame coordinate, $\omega_0$ is the central frequency, $k_0 = \omega_0/c$, and $c$ is the speed of light in vacuum. In vacuum, the transverse components of the envelope, $\mathbf{E}_\perp$, evolve according to

$$\left[2\left(ik_0 - \frac{\partial}{\partial \xi}\right)\frac{\partial}{\partial z} + \nabla_\perp^2\right]\mathbf{E}_\perp = 0. \quad (1)$$

The $\xi$-derivative, which is absent in the monochromatic paraxial equation, accounts for spatiotemporal delays and the frequency dependence of diffraction. We note that the approximations used to derive Eq. (1) break down for focusing geometries where the f-number, $f^{\#}$, ~4 or less. In these situations, a non-paraxial wave equation is required [20]. Accordingly, we limit this investigation to $f^{\#} > 6$. With a solution of Eq. (1), the approximate axial field can be found from $\nabla \cdot \mathbf{E} \approx 0$. For the parameters considered here $\mathbf{E}_0 \approx \mathbf{E}_\perp$.

The initial condition of Eq. (1), $\mathbf{E}_\perp(\mathbf{r},\xi,0)$, represents the envelope just after passage through the chromatic lens. For specificity, we use a diffractive lens, which produces a longitudinally extended chromatic focus while minimizing other aberrations [15]. A diffractive lens has a radially varying groove density $G = k_0 r / 2\pi f$, where $f$ is the focal length for the wavenumber $k_0$. This imparts a phase $\phi_{DL} = -k_0 r^2 / 2f$ to the pulse. Defining the envelope just before the diffractive lens as $\mathbf{E}_l(\mathbf{r},\xi,0)$, we have $\mathbf{E}_\perp(\mathbf{r},\xi,0) = \mathbf{E}_l(\mathbf{r},\xi,0)e^{-ik_0 r^2/2f}$.

The diffractive lens phase, $\phi_{DL}$, while similar to that imparted by an achromatic lens, differs in several important features. First, only the central frequency, $\omega_0$, will focus at the location $z \approx f$. The lower and higher frequency components will focus earlier and later respectively, an essential element of the flying focus. Second, the diffractive lens does not pre-compensate for the path length difference between 'rays' starting at the transverse edge of the pulse and those starting at the center. This causes the pulse to acquire an intensity curvature as it approaches focus similar to that discussed in Ref. [21]. From simple geometric considerations, the temporal delay as a function of radius in the diffractive lens plane is $\delta\xi(r) = (f^2 + r^2)^{1/2} - f$. The delay increases with decreasing f-numbers and causes a larger relative distortion for a short pulse than for a long pulse.

Given $\mathbf{E}_\perp(\mathbf{r},\xi,0)$, one can calculate the envelope at any axial location using the integral solution of Eq. (1) in the spectral domain (i.e. the modified Fresnel integral):

$$\hat{\mathbf{E}}_\perp(\mathbf{r},\delta\omega,z) = \frac{k_\omega}{2\pi i z} \int \exp\left[\frac{ik_\omega}{2z}(\mathbf{r}-\mathbf{r}')^2\right] \hat{\mathbf{E}}_\perp(\mathbf{r}',\delta\omega,z) d\mathbf{r}', \quad (2)$$

where the caret denotes a Fourier transform with respect to $\xi$ with conjugate variable $\delta\omega/c$ and $k_\omega = k_0 + \delta\omega/c$. Equivalently, one can perform the $\xi$-domain integral

$$\mathbf{E}_\perp(\mathbf{r},\xi,z) = -\frac{(ik_0 - \partial_\xi)}{2\pi z} \int \exp\left[\frac{ik_0}{2z}(\mathbf{r}-\mathbf{r}')^2\right] \mathbf{E}_\perp(\mathbf{r}',\xi - \tfrac{|\mathbf{r}-\mathbf{r}'|^2}{2z},0) d\mathbf{r}'. \quad (3)$$

Equation (3) can be interpreted as follows. The field in the transverse plane at any $z$ results from the superposition of light emitted from a collection of point sources in the plane $z = 0$. The first factor in the integrand (the propagator) accounts for the diffraction of the light emitted from the point sources. The initial condition $\mathbf{E}_\perp(\mathbf{r}',\xi,0)$ determines their amplitudes, phases, and time-dependence. The translation of the $\xi$ argument by $|\mathbf{r}-\mathbf{r}'|/2z$ accounts for the difference in time it takes light from point sources displaced transversely in the $z = 0$ plane to reach a particular location in the $z = z$ plane.

In the following, we will analyze Eqs. (2) and (3) for a FF pulse with a transverse Gaussian profile and a quadratic spectral phase (i.e. a chirp). Specifically,

$$\hat{\mathbf{E}}_\perp(\mathbf{r},\delta\omega,0) = \hat{\mathbf{E}}_a \exp\left[-\frac{r^2}{w_0^2} + \frac{1}{4}i\eta\tau^2\delta\omega^2 - \frac{ik_0 r^2}{2f}\right], \quad (4)$$

where $w_0$ is the initial spot size, $\tau$ is a measure of the transform-limited pulse duration, $\eta$ is the chirp parameter, and $\hat{\mathbf{E}}_a = \hat{\mathbf{E}}_a(\tfrac{1}{2}\tau\delta\omega)$ is the real, frequency-dependent spectral amplitude. While a different transverse profile may better represent a particular laser system, the Gaussian greatly simplifies analytic calculations using Eqs. (3) and (4), and illustrates the salient physics. The simulations in Section V will consider a different transverse profile.

### III. Analysis of Flying Focus Pulse Propagation

Equation (2), with the initial condition in Eq. (4), admits exact solutions. Each frequency component undergoes standard Gaussian optics diffraction with a slight modification. To obtain the spot size, curvature, Guoy phase or amplitude for each frequency, one simply applies the transformation $z \to (k_0/k_\omega)z$ to the monochromatic result. Of particular note is the spotsize:

$$w_\omega = w_0 \left[ \left( \frac{2z}{k_\omega w_0^2} \right)^2 + \left( 1 - \frac{k_0 z}{k_\omega f} \right)^2 \right]^{1/2} . \quad (5)$$

Equation (5) shows that each frequency has a different focal length $z_\omega \simeq (k_\omega/k_0)f$. For a power spectrum of width $\sim 4/\tau$, this spreads the focal region over a distance $\sim 4(\omega_0\tau)^{-1}f$. The effective focal length, $(k_\omega/k_0)f$, and angular width, $\sim 1/k_\omega w_0$, increase and decrease with frequency, respectively. The two exactly offset such that each frequency has the same minimum spot size—the diffraction-limited spot of the central wavenumber, $w_{min} = (2z_\omega/k_\omega w_0) = (2f/k_0 w_0)$.

In addition to focusing at different locations, each frequency can focus at a different time. The focal time for each frequency, $t_\omega$, consists of two contributions. The first results from the chromatic focusing of the diffractive lens and is simply the focal length of the frequency divided by the speed of light, $z_\omega/c$. The second results from the chirp and corresponds to the relative time of the frequency within the pulse. Denoting the spectral phase as $\phi_s$, the relative time is $\partial_{\delta\omega}\phi_s = \tfrac{1}{2}\eta\tau^2\delta\omega$. Using $\delta\omega = \omega_0(z_\omega/f - 1)$ and summing the two contributions provides $ct_\omega = z_\omega[1+\eta c\tau^2\omega_0/2f] - \tfrac{1}{2}\eta c\tau^2\omega_0$. Upon taking the derivative with respect to $t_\omega$, we arrive at the focal velocity

$$\frac{v_f}{c} = \left(\frac{2f}{\eta c\tau^2\omega_0}\right)\left(1+\frac{2f}{\eta c\tau^2\omega_0}\right)^{-1}. \quad (6)$$

The chromatic focusing of the diffractive lens and the temporal separation of frequencies results in a focal position that moves in time with the velocity $v_f$. An alternative, approximate derivation of Eq. (6) appears in Appendix A.

Figure (2) displays the focal velocity as a function of chirp and pulse duration for the parameters found in Table I. For a positive chirp, the red-shifted frequencies lead the blue-shifted frequencies in time. Because the red frequencies focus closer to the lens, they will necessarily focus earlier in time than the blue (see Fig. 1). The peak intensity, therefore, moves with positive velocity from the red to the blue focal points. As the chirp increases, the red and blue frequencies become more separated in time. This leads to an increase in time between their foci and a decrease in the focal velocity. For negative chirp, the blue-shifted frequencies lead the red. Depending on the value of the chirp, the blue frequencies can focus later or earlier in time than the red frequencies. The crossover point occurs when the time separation (multiplied by c) of the frequencies within the pulse equals the separation of their focal points, i.e. $\eta = -2f/c\omega_0\tau^2$. At the crossover point, all of the frequencies focus simultaneously producing a line focus. This is similar to an axicon "focus," with two important exceptions: all of the light comes to the line focus simultaneously, and the focal spot is that of a standard lens. Because of the crossover, negative chirps allow a wider range of focal velocities and provide greater versatility.

To aid in the conceptualization of a negative focal velocity, Fig. (3) displays intensity isocontours of a FF pulse with $v_f = -c/3$. The figure comprises three snapshots of the intensity in the laboratory frame, with time increasing from top to bottom. The spatial advance of the green and blue contours with time show that the pulse energy generally propagates in the forward direction (left to right). The peak intensity, on other hand, travels backwards at $v_f = -c/3$ as traced out by the solid grey line. This illustrates the principal feature of the flying focus: *the peak intensity travels at* $v_f$, *not the group velocity*. The simulations used to generate the isocontours are described in Section V and Appendix C.

Unfortunately, exact analytic solutions to Eq. (3) exist only for either trivial, e.g. a monochromatic Gaussian beam, or limiting cases. In Appendix B, for instance, we present an exact, but unwieldy, expression for the on-axis field of the flying focus pulse. For the remainder of the manuscript, we make the simplifying approximation of large chirp, $|\eta| \gg 1$. For clarity of presentation in this section, we make an additional simplifying approximation that is valid over the parameter range of interest: $4f^{\#}c\tau|\eta|/w_0 \gg 1$. Physically, this condition implies that the temporal delays for different path lengths from the diffractive lens to the focus are small compared to the chirped pulse duration. We note that the simulations, presented below, do not make this approximation, but validate its use here.

A large chirp admits use of the stationary phase approximation when performing the Fourier transform to find the $\xi$-domain profile from Eq. (4). Specifically, we have $\mathbf{E}_\perp(0,\xi,0) \approx \alpha \hat{\mathbf{E}}_a(\xi/cT)\exp[-i\eta(\xi/cT)^2]$, where $\alpha = (1+i)(\eta/2\pi T^2)^{1/2}$ and $T = \eta\tau$. For a super (or regular) Gaussian power spectrum of order $g$, Eq. (3) reduces to

$$\mathbf{E}_\perp(\mathbf{r},\xi,z) = \frac{k_0 \mathbf{F}(\xi)}{2\pi i \tilde{z}} \int \exp\left[\frac{ik_0}{2\tilde{z}}(\mathbf{r}-\mathbf{r}')^2 - \frac{ik_0 r'^2}{2f} - \frac{r'^2}{w_0^2}\right] d\mathbf{r}', \quad (7)$$

where $\mathbf{F}(\xi) = \mathbf{E}_a \exp[-(\xi/cT)^g - i\eta(\xi/cT)^2]$, $\mathbf{E}_a$ is a constant amplitude, and

$$\tilde{z} = \left(1 + \frac{2\eta\xi}{\omega_0 cT^2}\right)^{-1} z. \quad (8)$$

The integrand in Eq. (7) is nearly identical to that of a monochromatic Gaussian beam, but with the propagation distance scaled by a $\xi$-dependent factor: $z \to (1 + 2\eta\xi/\omega_0 cT^2)^{-1} z$. Expressions for the curvature and Guoy phases can be obtained by evaluating the monochromatic Gaussian beam results at $\tilde{z}$ instead of $z$. The resulting spot size $w = w_0[(\tilde{z}/Z_R)^2 + (1-\tilde{z}/f)^2]^{1/2}$, where $Z_R = k_0 w_0^2/2$, reproduces Eq. (5) with the frequency shift substitution discussed in Appendix A.

By writing Eq. (7) in terms of the scaled axial distance, $\tilde{z}$, an important property emerges: within the focal region, the envelope, and therefore the intensity, has a *self-similar form* that depends on $\xi$ and $z$ only in the combination $(1+2\eta\xi/\omega_0 cT^2)^{-1}z$. The intensity peak propagates with a velocity $v_f$ and a near-constant shape over a distance $\sim 4f(\omega_0\tau)^{-1}$—a distance that can readily exceed the Rayleigh length of the diffraction-limited spot. On axis, the intensity peak takes the simple form $I = \frac{1}{2}\varepsilon_0 cE_a^2(w_0/w)^2$, with temporal and spatial widths $\Delta\xi = 2\eta(c\tau w_0^{-1})^2 f$ and $\Delta z = (v_f/c)\Delta\xi = 2(1+2\eta f/\omega_0 cT^2)^{-1} f^2/Z_R$, respectively. Outside of the focal region, $|z-f| > 2f(\omega_0\tau)^{-1}$, $(\xi/cT)^2 > 1$, and $\xi$-dependence of the pulse amplitude, $|F(\xi)|$, breaks the self-similar form. A visual illustration of this will appear in the simulations presented below.

## IV. IWAV Model

As discussed above, the flying focus decouples the velocity of the intensity peak from the group velocity. By manipulating the speed of the peak, an ionization wave of arbitrary velocity (IWAV) can be produced. Aside from controlling the velocity of the ionization front, the ability to counterpropagate the intensity peak relative to the group velocity can mitigate the deleterious effects of ionization refraction.

A self-consistent model of IWAVs must include the propagation of the flying focus pulse, the ionization dynamics of the background gas, and the resulting plasma refraction and depletion of pulse energy. Extending Eq. (2) to include these effects, we have

$$\left[2\left(ik_0 - \frac{\partial}{\partial\xi}\right)\frac{\partial}{\partial z} + \nabla_\perp^2\right]\mathbf{E}_\perp = k_p^2\mathbf{E}_\perp - \mathbf{Q}. \quad (9)$$

where $k_p = (e^2 n_e/m_e c^2\varepsilon_0)^{1/2}$, $n_e$ is the free electron density, $e$ the fundamental unit of charge, $m_e$ the electron mass, and $\mathbf{Q}$ a function accounting for depletion, which we return to below. The FF pulses of interest here propagate through a tenuous, singly ionizable gas, such that $(k_p/k_0)^4 \ll 1$ consistent with the approximations used to derive Eq. (10).

At every spatial location, the gas starts un-ionized. Optical field ionization occurs throughout the pulse freeing electrons from the gas atoms. Once freed, the electrons can multiply by collisionally ionizing additional gas atoms. Eventually, the electrons recombine through radiative or three-body mechanisms. Noting that the ion density $n_i = n_e$ in a singly ionizable gas, the electron density evolves according to

$$c\partial_\xi n_e = \nu_{FI} n_g + \alpha_{CI} n_e n_g - \alpha_R n_e^2 - \beta_{3B} n_e^3, \quad (10)$$

where $n_g = n_{g0} - n_e$ and $n_{g0}$ is the initial gas density. The field ionization rate, $\nu_{FI}$, depends on the local value of the envelope, $\mathbf{E}_\perp$, while the rates for collisional ionization, $\alpha_{CI}$, radiative recombination, $\alpha_R$, and three-body recombination, $\beta_{3B}$, depend on the local electron temperature, $T_e$. In the simulations presented below, the cycle-averaged Ammosov-Delone-Krainov (ADK) rate is used for $\nu_{FI}$ [22], while the other rates are calculated as in Ref. [23].

Electron-ion collisions convert the laser pulse energy to electron thermal energy (i.e. inverse bremsstrahlung heating). For short, high-intensity laser pulses, the heating can occur rapidly, resulting in the loss of local thermodynamic equilibrium between the electrons and ions. We denote the electron and ion thermal energy densities as $\Theta_e = \tfrac{3}{2} n_e k_b T_e$ and $\Theta_i = \tfrac{3}{2} n_e k_b T_i$ respectively, where $T_i$ is the ion temperature. Their modifications are described by

$$c\partial_\xi \Theta_e = \frac{2\omega_0^2}{(\omega_0^2 + \nu_{ei}^2)} \nu_{ei} n_e U_p - \alpha_{CI} n_e n_g U_I - 3\frac{m_e}{m_i} \nu_{ei} n_e k_b (T_e - T_i) \quad (11a)$$

$$c\partial_\xi \Theta_i = 3\frac{m_e}{m_i} n_e \nu_{ei} k_b (T_e - T_i), \quad (11b)$$

where $\nu_{ei}$ is the standard electron-ion collision frequency [23], $U_p = m_e (e|E_\perp|/2m_e\omega_0)^2$ is the cycle-averaged kinetic (ponderomotive) energy of an electron within the pulse, $U_I$ the ionization potential, and $m_i$ the ion mass. In order, the terms on the right hand side (RHS) of Eq. (11a) correspond to inverse bremsstrahlung heating, the energy expended by free electrons during collisional ionization, and electron-ion thermal relaxation.

Field ionization and inverse bremsstrahlung deplete the pulse energy as it propagates through the gas or plasma. We express $\mathbf{Q}$ as a sum of the two contributions: $\mathbf{Q} = \mathbf{Q}_{FI} + \mathbf{Q}_{IB}$. For every electron the pulse frees from a gas atom, it must expend the ionization energy, $U_I$. This loss is captured by

$$\mathbf{Q}_{FI} = \left( ik_0 - \frac{\partial}{\partial \xi} \right) \frac{U_I \nu_{FI} n_g}{I} \mathbf{E}_\perp, \quad (12)$$

where $I = \frac{1}{2} \varepsilon_0 c |E_\perp|^2$ is the intensity. The inverse bremsstrahlung contribution,

$$\mathbf{Q}_{IB} = \left( ik_0 - \frac{\partial}{\partial \xi} \right) \frac{\omega_p^2}{\omega_0^2 + \nu_{ei}^2} c^{-1} \nu_{ei} \mathbf{E}_\perp, \quad (13)$$

balances the energy gained by the electrons through the first term in Eq. (11a).

**V. FF and IWAV simulations**

To demonstrate the propagation of the FF pulse and IWAV formation, we perform simulations that solve Eqs. (9-13) for the parameters displayed in Table I. Appendix C provides an overview of the method. The simulated pulses model the output from the frontend of the Multi-Terawatt laser (MTW) at the University of Rochester, Laboratory for Laser Energetics. MTW uses an optical parametric chirped pulse amplifier (OPCPA) frontend to pump a 100 J-class Nd:glass final amplifier. The OPCPA produces linearly polarized pulses with a central wavelength $\lambda_0 = 2\pi / k_0 = 1.054$ $\mu$m and 9.2 nm of bandwidth, full width at half maximum. The pulses have a nearly flat power spectrum over the entire bandwidth with a transform-limited duration of $\sim 1.6\tau \sim 370$ fs. Numerically, we implement the flat power spectrum as a super-Gaussian of order eight (SG8). The initial transverse profile, also an SG8, has a spot size, $w_0$, equal to the radius of the diffractive lens used for the experiments in Ref. [15]. Explicitly, $\hat{\mathbf{E}}_l \propto \exp[-(rw_0^{-1})^8 - (\frac{1}{2}\tau\delta\omega)^8 + \frac{1}{4}i\eta\tau^2\delta\omega^2]$. As in Ref. [15], the diffractive lens has a focal length $f = 0.51$ m at $\lambda_0$. Unless otherwise stated, the peak power is fixed at 210 MW, which corresponds to energies ranging from 3 to 9 mJ depending on the chirp. The pulses propagate through either ½ atmosphere of $H_2$ ($n_{g0} = 1$ atm of H atoms) or, for reference, vacuum.

Figure (4) displays lineouts of the on-axis ($r = 0$) intensity and electron density as a function of time and space for $v_f = -c$, corresponding to the red circle in Fig. (2). To guide the eye, a white dashed line demarcating a trajectory moving at $-c$ has also been plotted. The FF pulse created a sharp ionization front, i.e. an IWAV, that moves backwards at $v_f$. The on-axis intensity in Fig. (4) appears to have a shorter duration than would be expected from the chirp and power spectrum. This apparent shortening results from plasma refraction. The ionization rate has a highly nonlinear dependence on the electric field. As a result, ionization predominately occurs near the peak of the pulse, localizing the plasma to the center of the spot. The corresponding refractive index has a sharp gradient that strongly refracts the back of the pulse. The repercussions of plasma refraction can be observed more clearly in Fig. (5). Figure (5) displays the spatiotemporal profile of the FF pulse at three different locations in the lab frame. The profiles in $H_2$ are shown above the grey line, and, for reference, the vacuum profiles are shown below. The refraction of the back of the pulse is apparent.

The insets in Fig. (5) magnify the region around the peak intensity in each frame. The region about the peak intensity looks almost identical in all three frames. As discussed above, the FF pulse has a self-similar profile when propagating in vacuum. Surprisingly, a self-similar structure persists in the presence of the ionization refraction, albeit slightly modified. Figure (6) shows lineouts of the on-axis intensity at the same locations. Within the focal region, the on-axis intensity, and resulting plasma, has nearly the same temporal profile. Note that most of the ionization occurs behind the pulse. At the pulse powers considered here, optical field ionization provides the initial seed electrons for collisional ionization. Throughout its duration, the pulse heats the electrons, which allows collisional ionization to persist after the pulse has passed.

While Fig. (5) clearly demonstrates an effect of ionization refraction, the disruption to propagation was relatively benign: by counterpropagating the focus with respect to the group velocity, the brunt of ionization refraction was avoided. Figure (7) illustrates the advantage of moving the laser pulse backwards with respect to the group velocity. The figure displays a comparison of IWAVs created by FF pulses with $v_f = -c/2$ and $v_f = c/2$ on the left and right, respectively. The two velocities correspond to the red and blue squares in Fig. (2). For $v_f = -c/2$, the blue-shifted frequencies focus earlier and further from the lens, initiating the ionization wave. As the red-shifted frequencies come to focus and ionize, they do so closer to the lens, behind (in space) the already generated plasma. This allows the peak intensity to propagate backwards relatively unimpeded. For $v_f = c/2$, the red-shifted frequencies focus earlier and closer to the lens. As the blue-shifted frequencies come to focus, they propagate through the plasma created by the red-shifted frequencies. This exacerbates the ionization refraction, leading to a drastically lower electron density with an intermittent, disjointed profile as observed on the right in Fig. (7). In both cases, the depletion of pulse energy is less than 4% and has little affect on the propagation.

**VI. Advanced FF and IWAV Techniques**

In the previous section, we considered a laser pulse with a quadratic spectral phase focused by a diffractive lens. The peak intensity of the pulse, and the resulting ionization front, travelled at a constant velocity determined by the coefficient of the spectral phase (i.e the chirp). By using a more complex spectral phase, the peak intensity can be made to travel with a dynamic velocity [16]. Following the procedure in Section III, the trajectory of the peak satisfies

$$ct = z + \frac{f}{\omega_0} \partial_z \phi_s, \quad (14)$$

where we have used $\delta\omega = \omega_0(z/f - 1)$ to write $\phi_s$ in terms of $z$. For a desired $z(t)$, one can invert Eq. (14) to find $\phi_s$. As a practical example, one could accelerate or decelerate the speed of an IWAV to modulate the THz emission in the two-color optical Cherenkov mechanism [2,24].

One can also control the spatiotemporal properties of the IWAV by modifying the power spectrum. Recall that each frequency focuses at a different location on the optical axis and, with the exception of the line focus, a different time. As a result, the spectral amplitude of each frequency determines the peak intensity at each location and time. By adjusting the relative amplitude of each frequency, the trajectory of the peak intensity can be controlled independently of the focal velocity and chirp.

Here we consider the simplest case of a line focus, where all of the frequencies focus simultaneously at different locations on the optical axis. Because the diffractive lens maps frequencies to axial locations, one can ramp or mask the power spectrum to vary the intensity, and hence the ionization and heating, along the optical axis. To illustrate this, we apply a periodic modulation to the power spectrum: $\hat{E}_a \propto [1+\cos(\frac{1}{2}N\pi\tau\delta\omega)]^2 \exp[-(\frac{1}{2}\tau\delta\omega)^8]$. When focused, the spectrum produces an intensity profile that oscillates along the optical axis with a period $4f/N\tau\omega_0$. Figure (8) shows the on-axis intensity and electron density resulting from a modulated power spectrum with $N=6$. The peak power and energy of the pulse were 470 MW and 5 mJ, respectively. Similar profiles, created with axicons, have been used to enhance a variety of plasma-based applications, including THz radiation, betatron X-rays, and electron acceleration [25-27]. The FF, however, has several advantages over an axicon: (1) at each longitudinal location, the FF has a minimum spot size nearly equal to that of an ideal lens, (2) for a flat power spectrum, the peak, on-axis intensity of the FF does not vary within the focal region, and (3) the FF largely avoids the plasma refraction that limits the intensity achievable by an axicon.

**VII. Summary and Conclusions**

We have examined the formation of ionization waves of arbitrary velocity (IWAVs) by "flying focus" pulses. A dynamic, or flying, focus occurs when a chromatic lens focuses a laser pulse with a nonlinear spectral phase. In a region about the focus, the peak intensity of the laser pulse follows a time-dependent trajectory along the optical axis. Generally, the trajectory depends on the details of the spectral phase. For the special case of a quadratic spectral phase (i.e. a chirp), the peak intensity travels with a constant velocity that is distinct from the group velocity of the pulse. This distinct "focal velocity" can be tuned through the chirp such that the peak co- or counterpropagates with respect to the optical axis. In either case, the peak propagates self-similarly through the focal region with a near-constant spatiotemporal profile. The length of the region, determined by the bandwidth of the pulse and focal length of the lens, can readily exceed the Rayleigh range of the diffraction-limited spot.

In a medium, the flying focus pulse field-ionizes the constituent atoms and heats the resulting electrons along the trajectory of its peak intensity. The leading edge of the ionization, both field and collisional, follows the trajectory of the peak intensity. Thus by adjusting the chirp, an ionization front can be made to travel at an arbitrary velocity along the optical axis. This property has the potential to enable a wide range of applications currently limited or precluded by an inability to phase-match at the laser pulse group velocity, such as THz generation, high harmonic generation, and photon acceleration. From a practical standpoint, a negative focal velocity pulse can create a clean ionization front relatively unimpeded by plasma refraction. As a more exotic example, one can mask or ramp the power spectrum to further control the propagation of the FF pulse and tailor the plasma formation, for instance creating modulated plasma density profiles.

**Acknowledgements**

The authors would like to thank T. Kessler, A. Solodov, J. Shaw, A. Sefkow, J. Vieira, N. Vafaei-Najafabadi, S. Bucht, and A. Davies, for fruitful discussions.

This work was supported by the U.S. Department of Energy Office of Fusion Energy Sciences under contract No. DE-SC0016253, Department of Energy under Cooperative Agreement No. DE-NA0001944, the University of Rochester, and the New York State Energy Research and Development Authority.

**Appendix A: Alternative Derivation of the Focal Velocity**

Equation (6) accurately predicts the focal velocity for an arbitrary chirp [15,16]. For an alternative, rough derivation of the flying focus velocity, we consider the limit of large chirp. In this limit, the stretched pulse duration, $T$ (e.g. $T = (1+\eta^2)^{1/2}\tau$ for a Gaussian power spectrum), far exceeds the transform limited duration. We can then write the frequency shift in $z_\omega$ as its $\xi$-domain representation: $\delta\omega(\xi) \simeq 2\eta\xi_\omega / cT^2$. The focal location for each frequency becomes

$$\frac{z_\omega}{f} \simeq \left(1 + \frac{2\eta f}{c\omega_0 T^2}\right)^{-1} \left(1 + \frac{2\eta t_\omega}{\omega_0 T^2}\right), \quad (A1)$$

with the corresponding velocity $v_f \simeq (1 + 2\eta f / c\omega_0 T^2)^{-1} (2\eta f / \omega_0 T^2)$. For large chirp $T \approx \eta\tau$, and this expression reproduces Eq. (6).

**Appendix B: On-axis FF Solution**

In this appendix, we find an exact solution to Eq. (3) evaluated on the propagation axis, $\mathbf{r} = 0$, for a spatiotemporal Gaussian pulse shape. Using a Gaussian power spectrum and Fourier transforming Eq. (4) to the $\xi$-domain, we have

$$E_0(\mathbf{r},\xi,z) = E_a \exp\left[-\frac{r^2}{w_0^2} - \frac{ik_0 r^2}{2f} - \frac{\xi^2}{c^2\tau^2(1-i\eta)}\right], \quad (B1)$$

where $E_a$ is a constant amplitude. Inserting this expression into Eq. (3) and setting $\mathbf{r} = 0$, provides

$$E_0(0,\xi,z) = \frac{E_a(k_0 + i\partial_\xi)}{2\pi i z} F(\xi) \int \exp[-qr'^2 - pr'^4] d\mathbf{r}', \quad (B2)$$

where $F(\xi) = E_a \exp[-(1+i\eta)\xi^2 / c^2 T^2]$, $T = (1+\eta^2)^{1/2}\tau$, $p = (1+i\eta)/4c^2 T^2 z^2$, and

$$q = \frac{k_0}{2z}\left[\left(\frac{2z}{k_0 w_0^2} - \frac{2\xi}{k_0 c^2 T^2}\right) - i\left(1 - \frac{z}{f} + \frac{2\eta\xi}{k_0 c^2 T^2}\right)\right].$$

At this point, one may already discern that $(2w_0 z / k_0)|q|$ has the appearance of the spot size. Indeed, we recover our $\xi$-domain expression for the spot size from Section III when $|\eta| \gg 1$, i.e. when we drop the second term in $q$. Completing the integral in Eq. (B2) yields the envelope of the FF pulse evaluated on axis:

$$E_0(0,\xi,z) = \frac{\pi^{1/2} E_a}{4izp^{1/2}} (k_0 + i\partial_\xi) F(\xi) e^{q^2/4p} \text{erfc}\left(\frac{q}{2p^{1/2}}\right), \text{(B3)}$$

where erfc is the complimentary error function. As expected, Eq. (B3) recovers the monochromatic Gaussian optics result in the limit $T \to \infty$. Equation (B3) appears somewhat unwieldy, motivating our use of simplifying approximations and simulations in Sections III and V respectively.

**Appendix C: Discussion of Simulation Method**

Numerical solutions to Eq. (1) or Eqs. (10-14) can be obtained by applying standard pseudo-spectral split step methods [29]. In these methods, one makes many small advances in $z$, alternating between diffraction steps in Fourier space and nonlinear or refractive steps in real space. If initialized at the diffractive lens plane, these methods can come with great computational expense. One must resolve the largest wavenumber imparted by the diffractive lens (equivalent to resolving the spot size at focus), while making the transverse simulation domain large enough to contain the initial spot size. Denoting the transverse resolution by $\Delta x$ and the domain size as $N\Delta x$, these requirements can be expressed as $\Delta x < 2\pi k_0^{-1} f^\#$ and $\Delta x > 2w_0 / N$ respectively. The resulting number of grid points can be extremely large: $N \sim k_0 w_0 / \pi f^\#$. Consider 1 $\mu$m wavelength light focused through a 3.6 cm radius diffractive lens with an $f^\# = 7$. The bare minimum number of grid points—one cell in a vacuum spot—in a single transverse dimension would be $N \sim 9800$. In terms of Eqs. (2) and (4), the first requirement amounts to resolving the rapidly varying phase appearing in the kernel: $k_0 w_0 \Delta x / f < 1$.

We can avoid such an onerous computation by recognizing that (1) the dynamics of interest occur near focus and (2) near focus, the rapidly varying phase of the kernel is nearly cancelled by the phase applied by the diffractive lens. Using Eq. (2) to find the envelope near focus decouples the numerical grids at $z=0$ and $z \sim f$, and greatly relaxes the requirements on the number of grid points. Convergence of the integral requires $k_0 w_0 \Delta x' | L^{-1} - f^{-1} | < 1$, where $\Delta x'$ is the transverse resolution in the diffractive lens plane. This significantly reduces the number of grid points required in one transverse direction $N' \sim |1 - L/f| (k_0 w_0 / \pi f^{\#})$. Meanwhile, the transverse domain in the $z$-plane needs only to contain the spot size at $z$, and can have whatever resolution is desired (for a rough estimate of the domain size one can simply use the spot size predicted by Gaussian optics). With these requirements met, the integral in Eq. (2) can be performed using standard numerical techniques, and the temporal profile found by performing a fast Fourier transform (FFT) with respect to $\delta\omega$.

In the simulations presented here, Eq. (2) is used to propagate the flying focus pulse from the diffractive lens to a location before focus where the intensity is just below the ionization threshold, $z = z_i$. As discussed above, this greatly relaxes the requirements on the numerical grid. From $z = z_i$ to $z = z_f$, where the intensity is above the ionization threshold, a pseudo-spectral split step method is used to solve Eq. (10).

Figure (3) was generated from the output of a 2D+t cylindrically symmetric simulation that used a discrete Hankel transform in the transverse direction. The numerical implementation of the Hankel transform was found to be highly dissipative, failing to conserve the energy of the pulse. Because Eq. (1) is linear, this was not an issue for examining FF pulse propagation in vacuum. However, the dissipation can cause spurious results when looking at the nonlinear problem of IWAV formation: the ionization rate depends strongly on the local intensity. To avoid this, the remaining figures, (4)-(8), were generated from the output of 2D+t Cartesian simulations that used FFTs.

| Pulse Parameters | Value |
| --- | --- |
| λ (μm) | 1.054 |
| Δλ (nm) | 9 |
| τ (fs) | 230 |
| P (MW) | 210 |
| Diffractive Lens Parameters | Value |
| $w_0$ (cm) | 3.6 |
| f (cm) | 51 |
| Gas Parameters | Value |
| $n_g$ (cm$^{-3}$) | $2.7 \times 10^{19}$ |
| $U_I$ (eV) | 13.6 |
| $T_0$ (eV) | .026 |

Table 1. Parameters for simulations

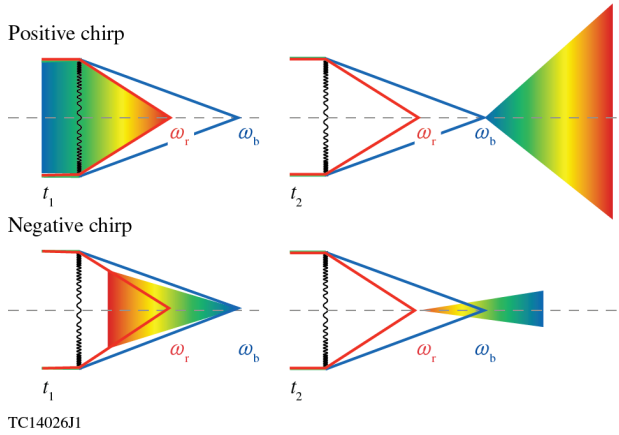

TC14026J1

Figure 1. Schematic of the flying focus. A positively (top row) or negatively (bottom row) chirped laser pulse passes through a diffractive lens. For the positive chirp, the red frequencies come to focus earlier than the blue, resulting in a positive focal velocity. For the depicted negative chirp, the blue frequencies come to focus earlier, resulting in a negative focal velocity.

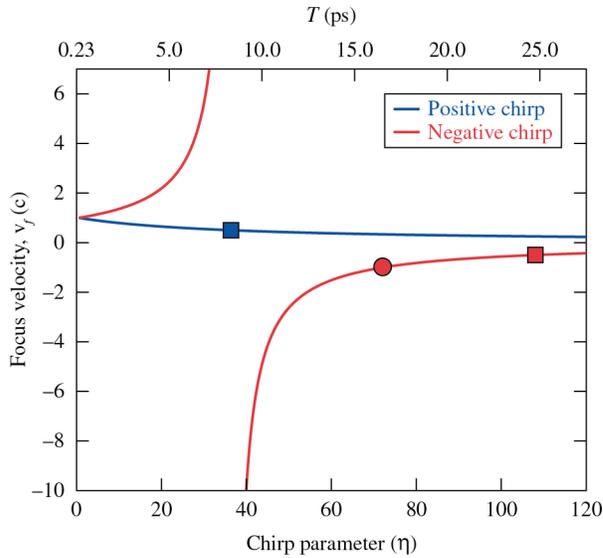

Figure 2. The focal velocity of the flying focus pulse as a function of chirp parameter and pulse duration for the parameters in Table I. A positively chirped pulse is limited to positive focal velocities, while a negatively chirped pulse can have either a positive or negative velocity.

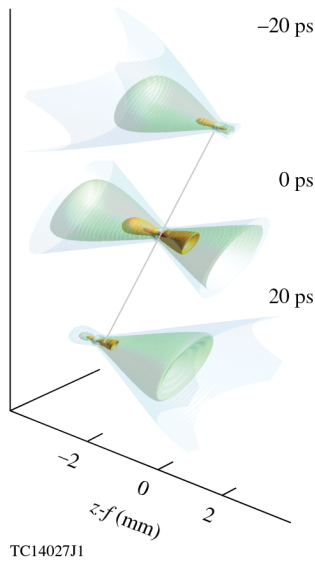

TC14027J1

Figure 3. Intensity isocontours of a flying focus pulse with $v_f = -c/3$ in the laboratory frame at three times, increasing from top to bottom. The peak intensity, represented by the dark orange contours, counterpropagates with respect to the optical axis.

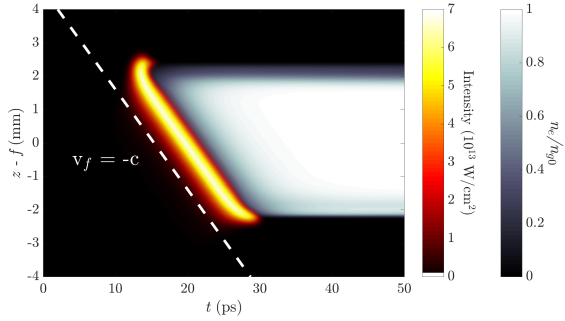

Figure 4. Intensity and electron density as a function of axial location and time for a flying focus pulse with $v_f = -c$. The pulse creates a sharp ionization front that also travels at $-c$.

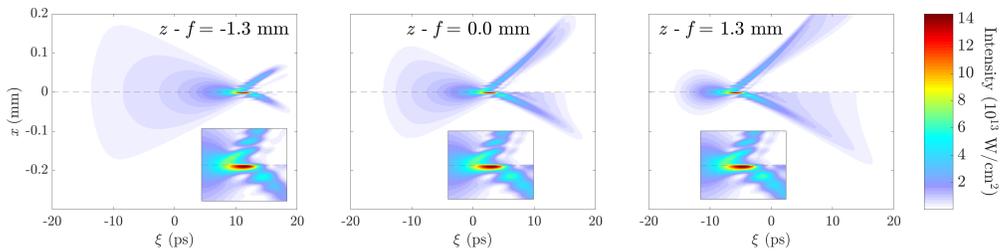

Figure 5. Spatiotemporal intensity profiles of a flying pulse with $v_f = -c$ at three axial locations. The profile in $H_2$ is plotted above the dashed line, and, to emphasize the effect of plasma refraction, the vacuum profiles are plotted below. The insets illustrates that, near peak intensity, a self-similar structure persists in both $H_2$ and vacuum as the flying focus traverses the focal region.

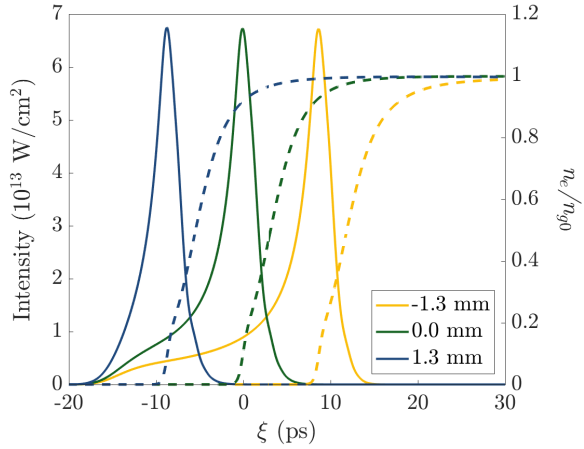

Figure 6. The on-axis intensity of a flying pulse with $v_f = -c$ at three axial locations. The persistence of a self-similar intensity profile, even with ionization dynamics, is evident.

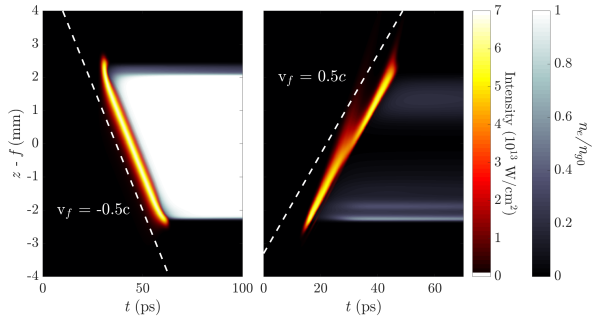

Figure 7. Intensity and electron density as a function of axial location and time for flying focus pulses with $v_f = -c/2$ and $v_f = c/2$ on the left and right respectively. The $v_f = -c/2$ pulse creates a sharp, clean ionization front, largely avoiding plasma refraction. The $v_f = c/2$ pulse suffers significant plasma refraction, producing an intermittent, disjointed ionization front.

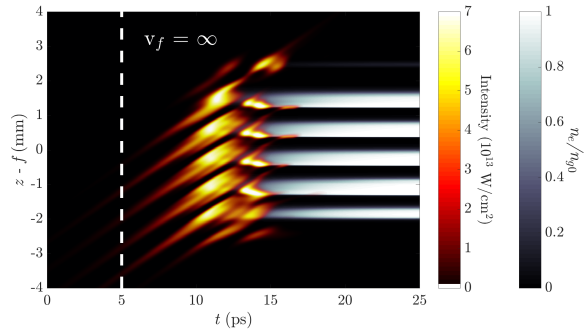

Figure 8. Intensity and electron density as a function of axial location and time for a flying focus pulse with a modulated power spectrum and $v_f = \infty$, i.e. a simultaneous line focus. The diffractive lens has mapped the modulation in the power spectrum to a spatial intensity modulation. A modulated plasma density profile results.